\documentclass[runningheads]{llncs}
\usepackage[T1]{fontenc}
\usepackage{graphicx}
\usepackage{amsfonts} 
\usepackage{amsmath}
\usepackage{multirow}
\usepackage{tabularx}
\usepackage{algorithm}
\usepackage{algpseudocode}
\usepackage{svg}
\usepackage{hyperref}

\begin{document}
\title{Reliable Multi-View Learning with Conformal Prediction for Aortic Stenosis Classification in Echocardiography}
\titlerunning{Reliable Multi-View AS Classification with Conformal Prediction}
\author{Ang Nan Gu\inst{1} \and Michael Tsang\inst{2} \and Hooman Vaseli\inst{1} \and Teresa Tsang\inst{2} \and Purang Abolmaesumi\inst{1}}
%
\authorrunning{A. Gu et al.}
%
\institute{Department of Electrical and Computer Engineering, University of British Columbia, Vancouver, Canada \\
\email{guangnan@ece.ubc.ca}\\
 \and
Division of Cardiology, Vancouver General Hospital, Vancouver, Canada
\footnote{T. Tsang and P. Abolmaesumi are joint senior authors.}
}
\maketitle              
\begin{abstract}
The fundamental problem with ultrasound-guided diagnosis is that the acquired images are often 2-D cross-sections of a 3-D anatomy, potentially missing important anatomical details. This limitation leads to challenges in ultrasound echocardiography, such as poor visualization of heart valves or foreshortening of ventricles. 
Clinicians must interpret these images with inherent uncertainty, a nuance absent in machine learning’s one-hot labels.
We propose Re-Training for Uncertainty (RT4U), a data-centric method to introduce uncertainty to weakly informative inputs in the training set. This simple approach can be incorporated to existing state-of-the-art aortic stenosis classification methods to further improve their accuracy. When combined with conformal prediction techniques, RT4U can yield adaptively sized prediction sets which are guaranteed to contain the ground truth class to a high accuracy. 
We validate the effectiveness of RT4U on three diverse datasets: a public (TMED-2) and a private AS dataset, along with a CIFAR-10-derived toy dataset. 
Results show improvement on all the datasets.
Our source code is publicly available at: \url{https://github.com/an-michaelg/RT4U}

\end{abstract}
%
%
\section{Introduction}
\subsubsection{Clinical motivation.} 
Aortic Stenosis (AS) is a heart valve disease in which the restricted motion of the aortic valve leaflets (AVL) impairs blood circulation. 
AS is highly prevalent in older populations \cite{ancona2020epidemiology}, posing a growing societal burden with global aging trends. Early detection, monitoring, and timely intervention are crucial due to the low survival rate among untreated AS patients \cite{strange2019poor}. AS diagnosis typically involves echocardiography (echo), with the decision on disease severity relying mostly on blood flow measurements from spectral Doppler \cite{bonow2006acc}. 
However, reliance on Doppler limits screening for AS to more specialized echo laboratories that have access to high-end cart-based echo machines. Acquiring spectral Doppler specifically requires very high level of expertise in echo imaging.
Recent clinical~\cite{abe2021screening,gulivc2016pocket,nemchyna2021validity} and machine learning \cite{ahmadi2023transformer,dai2023identifying,AS_Tom,guo2021predicting,holste2023severe,wesslerAutomatedAS2023} works show that AS classification with B-mode echo can achieve high-sensitivity and specificity, which may pave the way for developing screening tools that can be used by less experienced point-of-care practitioners. This work focuses on improving the automated screening of AS with B-mode echo input. 

\subsubsection{Challenges in interpreting echo exam data.} 
Echocardiograms are 2-D (+time) visualizations of 3-D (+time) patient anatomy, captured by human sonographers. However, variations in patient size and operator skill can result in echo views that lack sufficient diagnostic information \cite{chamsi2017handheld}. 
Directly training a model to predict AS severity from any cross-sectional view can be misleading due to limited AVL visibility. This limitation holds true even for the parasternal long-axis (PLAX) and parasternal short-axis (PSAX) views, for which acquiring the correct angle to image the AVL requires operator skill.
Therefore, during inference, we would like the model to have robust uncertainty estimation (UE) to warn users and prevent failed predictions if the input is inadequate. To the best of our knowledge, simultaneously pursuing these two goals - learning with imperfect data and developing strong UE - is rarely done. 

\subsubsection{Conformal prediction.}
\begin{figure}[t]
    \centering
    \includegraphics[width=0.99\textwidth]{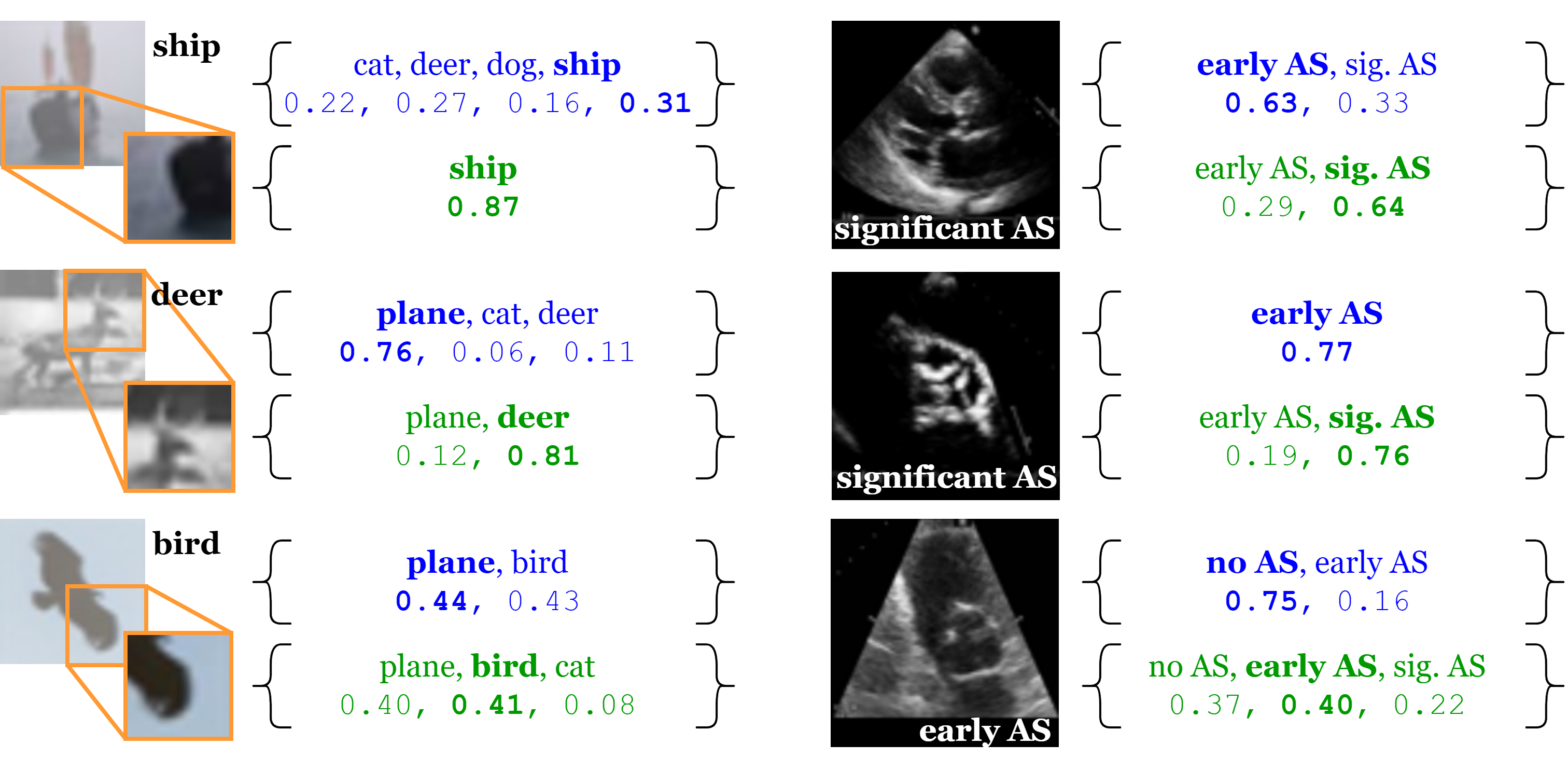}
    \caption{Conformal prediction returns a set $\{.\}$ of predictions. The set size correlates with the estimated uncertainty. Our \textbf{RT4U} method (green) enables better prediction and more adaptive set sizes compared to prior art (blue). Left: Inference on CIFAR-10 quadrants, orange box indicates the quadrant seen by the network. Right: Inference on echo cross-sections from the TMED-2 dataset. Ground truth is embedded in each echo image.}
    \label{fig:cpred}
\end{figure}

Neural networks predictions are often overconfident \cite{guo2017calibration}. Temperature scaling can mitigate this effect, but does not have a rigorous connection between the validation and test set performance. In contrast, Conformal Prediction (CP) is a way of transforming class predictions $p(y|x) \in [0,1]^K$ 
to prediction sets $C(x) \subset \{1, ..., K\}$ (Fig. \ref{fig:cpred}), where $K$ is the number of classes. Crucially, the sets are guaranteed to ``cover'' the ground truth to some user-specified probability \cite{angelopoulos2021intro} (see section \ref{sec:conformal_background}).

\subsubsection{Contributions.}
Inspired by a technique for learning with noisy labels \cite{chen2021seal}, we frame the problem as ``learning with noisy inputs''. The disparity between inputs affected by some inherent ambiguity (e.g., from poor image quality or cross-sectioning) and one-hot labels can be viewed as noise. Given that these labels fail to capture the ambiguity a clinician might encounter when analyzing the echo data, they should not be one-hot encoded. This type of noise, while uncommon in camera images, is ubiquitous in echo imaging due to the images being 2-D visualizations of 3-D anatomy. 
We propose Re-Training for Uncertainty (RT4U), a training method that uses pseudo-labels to mitigate overfitting to noisy inputs.
RT4U is a model-agnostic technique without hyperparameters, and it can be applied to existing approaches for AS classification with minimal overhead. We show that RT4U leads to improvements both top-1 accuracy and reduces prediction over-confidence.

\section{Background and Related Works}
\textbf{Automated detection of aortic stenosis.}
Deep learning approaches have been proposed for AS severity classification from 2-D echo. Huang et al.~\cite{huang2021tmed1} used images from multiple standard-plane views, and predicted both the view and the severity. Dai et al. \cite{dai2023identifying} and Holste et al.~\cite{holste2023severe} separated severe AS cases from other cases with PLAX view videos as the input. Ginsberg et al. \cite{AS_Tom} proposed an ordinal regression based approach. Ahmadi et al.~\cite{ahmadi2023transformer} used an attention-based architecture to identify key video frames for classification. Vaseli et al.~\cite{vaseli2023protoasnet} used an attention-based framework to identify important spatio-temporal components of the video. However, all the aforementioned works utilized one-hot labels. Secondly, none incorporated conformal prediction techniques to improve prediction reliability.
\newline
\newline
\textbf{Conformal prediction.}
\label{sec:conformal_background}
Conformal prediction originated from Vovk et al. \cite{vovk2005algorithmic}. 
Recent advancements such as LABEL \cite{sadinle2019label}, APS \cite{Romano2020aps}, and RAPS \cite{angelopoulos2022raps} have introduced CP algorithms with lower and more adaptive set sizes. 
In the medical domain, Lu et al. \cite{lu2022groupaps} validated CP on clinical data and extended RAPS to guarantee coverage for specific patient subgroups. Lu et al. \cite{lu2022improving} extended APS to ordinal regression problems. Stutz et al. \cite{stutz2023conformal} extended CP principles to data with multi-rater labels. CP principles have also been applied to skin disease \cite{zhang2023rrcp} and histopathology images \cite{wieslander2020wsi}.

Mathematical results from \cite{vovk2005algorithmic} show sets produced by CP can be guaranteed to a user-defined coverage error rate $\alpha$, e.g., $0.1$. 
In essence, prediction sets satisfy the following equation, also known as the ``Marginal Coverage Property'':
\begin{equation}\label{eqn:bound}
    1 - \alpha \leq \mathbb{P}(y_{test} \in C(x_{test})) \leq 1 - \alpha + \frac{1}{N+1},
\end{equation}
where $\mathbb{P}(y_{test} \in C(x_{test}))$ is the coverage accuracy, and $N$ is the number of datapoints used as calibration data. 
Higher coverage accuracy has a trade-off with higher prediction set size, denoted $|C(x)|$. Additionally, Eq.~(\ref{eqn:bound}) is not always satisfied due to the finite nature of the calibration set; the realistic coverage is distributed around $1 - \alpha$ \cite{vovk2012conditional}. 

\section{Methodology}

\subsection{Training dynamics with input-dependent noise}
\begin{figure}
    \centering
    \includegraphics[width=0.99\textwidth]{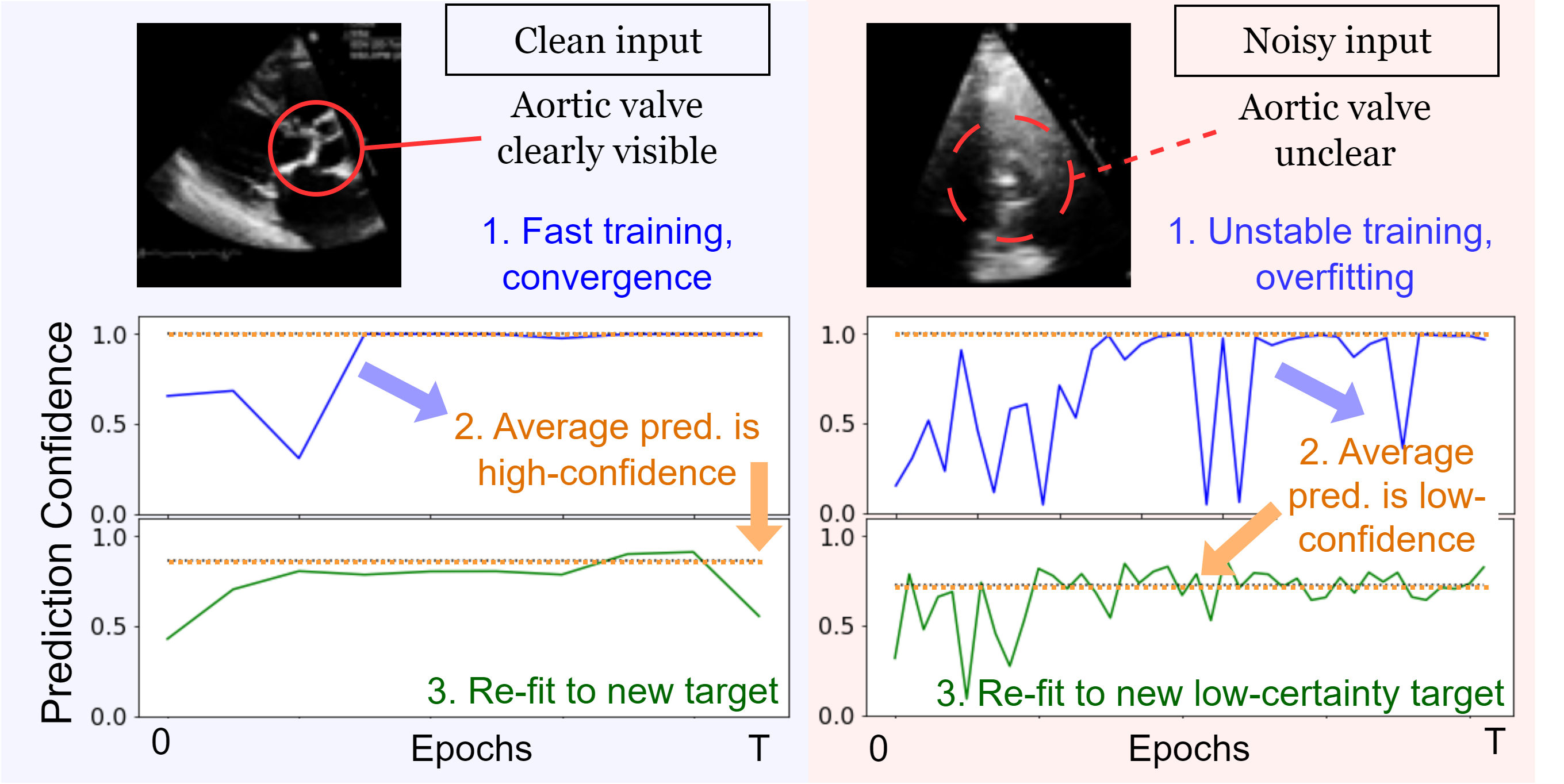}
    \caption{Demonstration of the RT4U algorithm in three steps. 1: When fitting on one-hot labels, the performance over different epochs vary depending on the clarity of the image (blue line). 2: Using the average of the past predictions, we form a pseudo-label (orange dotted line). 3: Fitting on the new pseudo-label improves stability and the model's uncertainty-awareness (green line).}
    \label{fig:evolution}
\end{figure}

Deep learning models can overfit complex forms of noise \cite{algan2021overfittingnoise}. This includes overfitting to non-informative ultrasound views, which can be considered as input-dependent noise. An observation by Chen et al. \cite{chen2021seal} shows that neural networks tend to fit simpler examples first. We confirm that this phenomenon also exists for ultrasound images (Fig. \ref{fig:evolution}). Consequently, the evolution of model's confidence during training can serve as a rough approximation of 
the `noise'' of a particular sample.
We can take advantage of this to create new pseudo-labels using the history of model predictions during training (Eq.~(\ref{eqn:pseudo})), where $T$ denotes the total number of epochs, $\sigma$ denotes softmax, and $\hat{f}(.)$ denotes the logits of the predictive model.
\begin{equation}\label{eqn:pseudo}
    y_i ' = \sum_{t=1}^{T} \sigma(\hat{f}(x_i)_t)
\end{equation}

\subsection{RT4U algorithm}
Our proposed RT4U algorithm can be applied to any iterative training approach.
First, we train the model with $\mathcal{D}_{tr} =\{X_{tr}, Y_{tr} \}$ for $T$ epochs, saving the prediction of training set $\hat{f}(x_i)_t \ \forall \ i \in X_{tr}, t \in T$. We form pseudo-labels $y'_{i} \ \forall \ x_i \in X_{tr}$ using Eq.~(\ref{eqn:pseudo}). Subsequently, we train the model from scratch with $(x_i, y'_i)$ pairs for $T$ epochs. This re-initialization is necessary for eliminating the impact of overfitting caused by training in the initial round. 
Subsequently, CP algorithms can be applied to the trained model to generate prediction sets. We opt for the LABEL algorithm \cite{sadinle2019label}. LABEL requires a calibration set $\mathcal{D}_{cal} = \{X_{cal}, Y_{cal} \}$ to find a conformal score quantile, $\hat{q}$. The $\hat{q}$ is then used as a threshold; classes with confidence exceeding it belongs to $C(x)$ for samples $x \in X_{test}$.

\section{Experiments}

\subsection{Simulated noise on CIFAR-10 dataset}
We first demonstrate the efficacy of our proposed model-agnostic methodology on a toy problem.
We generate quadrant-label pairs from images in the CIFAR-10 dataset, creating what we term \textbf{CIFAR-Q}. For each image-label pair $(\mathcal{I}_i, y_i)$, we train the network on the four image quadrants $(x_{i1}, y_i)$, $(x_{i2}, y_i), (x_{i3}, y_i), (x_{i4}, y_i)$ (see Fig.~\ref{fig:cpred}). CIFAR-Q simulates learning from potentially non-informative 2-D slices of 3-D objects, within a controlled environment constructed from a well-known dataset. We first split the CIFAR-10 training set randomly with a 95:5 ratio, then form $X_{tr}$ and $X_{val}$ from the quadrants of the splits. 

The setups compared are as follows: Vanilla training with cross-entropy loss (CE), CE with temperature scaling on $X_{val}$ (CE+Temp), CE with temperature scaling and RT4U training (CE+RT4U+Temp), Mean Absolute Error loss (MAE), and Active Negative Loss \cite{ye2024anl} (ANL) with hyperparameters $(\alpha, \beta, \delta) = (1, 1, 0)$. 

\subsection{Aortic stenosis classification using image and video data}
We integrate RT4U into state-of-the-art methods for AS classification. 

In the image domain, we compare with Huang et al. \cite{huang2021tmed1} on the public \textbf{TMED-2} dataset. This dataset contains 17270 images from 599 echo studies classified as no/early/significant AS. We use the training and validation splits from the provided DEV479 file, excluding studies lacking AS diagnosis labels.

In the video domain, we apply RT4U to enhance two existing approaches, R(2+1)D-18 \cite{AS_Tom} and ProtoASNet \cite{vaseli2023protoasnet}, on a private echo video dataset (\textbf{AS Private}), extracted from the echo database of a tertiary care hospital under ethics approval. 
Ultrasound videos were captured with Philips iE33, Vivid i, and Vivid E9 transducers. 
Each study was categorized as no/mild/moderate/severe AS based on Doppler echo guidelines \cite{bonow2006acc}, with moderate and severe grouped as significant. 
PLAX and PSAX videos were selected from each study by an experienced cardiologist. 
UI elements around the beam area were removed.
The dataset contains 5055 PLAX and 4062 PSAX videos from 2572 studies. We create training, validation and test sets using an 80:10:10 split, ensuring mutually exclusive sets of patients.
For R(2+1)D-18 \cite{AS_Tom}, we use the cross-entropy variant. For ProtoASNet, we use hyperparameters from \cite{vaseli2023protoasnet}.
All experiments are conducted on an NVIDIA TITAN V GPU. 
Further detail on experimental settings is provided in the supplementary material.


\subsection{Metrics}
We evaluate approaches with the following metrics at both the instance-level and study-level. ``Study'' refers to an echo study consisting of multiple images/videos, or an entire image consisting of four quadrants for CIFAR-Q. $BAcc$ and $BCov$ are class-balanced versions of the top-1 accuracy and coverage accuracy. We report the median $BCov$ and $C(x)$ from 100 random trials in order to confirm the marginal coverage property while accounting for randomness. 

\subsection{Decision aggregation}
We use different approaches for aggregating multiple instance-level predictions into a study-level prediction. For CIFAR-Q experiments and R(2+1)D \cite{AS_Tom}, we employ the method in Satop{\"a}{\"a} et al. \cite{satopaa2014combining} to average the instance-level predictions in the logit space (Eq.~(\ref{eqn:combine})), where $x_j, j \in [1, M]$ denote the $M$ instances within the same study.
\begin{equation}\label{eqn:combine}
    \mathbb{P}(y|x_1, x_2.. x_M) = \sigma(\sum_{j=1}^{M} \hat{f}(x_j))
\end{equation}
For Huang et al. \cite{huang2021tmed1}, we perform a weighted average based on the predicted view probability of PSAX or PLAX.
We approximate the view probability with a ResNet-18 model.
For ProtoASNet \cite{vaseli2023protoasnet}, we perform an average of the instance-level predictions $\in [0,1]^K$ with consideration of the additional ``uncertainty'' class that is introduced by the ProtoASNet architecture. These aggregation approaches are consistent with their original works.



\begin{table}[tb]
\centering
\caption{\textbf{Instance-level} test set evaluation. We aim to maximize $BAcc$, minimize $|C(x)|$, and maintain $BCov$ above the user-defined coverage, $1-\alpha$.}
\begin{tabular}{l|l|c|c|c}
\hline
Dataset                  & Method            & $ \ BAcc \uparrow \ $ & $ \ BCov \ $ & $ \ |C(x)| \downarrow \ $ \\ \hline \hline
\multirow{5}{*}{\begin{tabular}[c]{@{}l@{}}CIFAR-Q\\ $\alpha = 0.05$ \end{tabular}} 
                         & ANL \cite{ye2024anl}               & 0.734   & 1.000  & 10.0             \\
                         & MAE               & 0.735   & 0.950  & 2.90             \\
                         & CE                & 0.743   & 0.950  & 2.38             \\
                         & CE+Temp          & 0.743   & 0.950  & 2.37            \\
                         & CE+RT4U  & \textbf{0.765}   & 0.950  & \textbf{2.22}             \\
                         & CE+RT4U+Temp  & \textbf{0.765}   & 0.950  & \textbf{2.22}            \\ \hline
\multirow{4}{*}{\begin{tabular}[c]{@{}l@{}l@{}}AS\\ Private \\ $\alpha = 0.1$ \end{tabular}} 
                         & R(2+1)D \cite{AS_Tom} & 0.747 & 0.902 & \textbf{1.41}  \\
                         & R(2+1)D+RT4U      & \textbf{0.761}   & 0.902  & \textbf{1.41}              \\
                         & ProtoASNet \cite{vaseli2023protoasnet}       & 0.750   & 0.903  & 1.46            \\
                         & ProtoASNet+RT4U   & 0.755   & 0.904  & 1.46          \\ \hline
\multirow{2}{*}{\begin{tabular}[c]{@{}l@{}}TMED-2\\ $\alpha = 0.1$ \end{tabular}}  & Huang et al. \cite{huang2021tmed1}            & 0.610   & 0.907  & 2.05  \\
                         & Huang et al.+RT4U        & \textbf{0.661}   & 0.905  & \textbf{1.89}        \\ \hline
\end{tabular}
\label{tab:instance_level}
\end{table}

\begin{table}[tbh]
\centering
\caption{\textbf{Study-level} test set evaluation. We aim to maximize $BAcc$, minimize $|C(x)|$, and maintain $BCov$ above the user-defined coverage, $1-\alpha$.}
\begin{tabular}{l|l|c|c|c}
\hline
Dataset                  & Method            & $ \ BAcc \uparrow \ $ & $ \ BCov \ $ & $ \ |C(x)| \downarrow \ $ \\ \hline \hline
\multirow{5}{*}{\begin{tabular}[c]{@{}l@{}}CIFAR-Q\\ $\alpha = 0.05$ \end{tabular}} 
                         & ANL \cite{ye2024anl}     & 0.883   & 0.950  & 1.32   \\
                         & MAE               & 0.887   & 0.949  & 1.28              \\
                         & CE                & 0.901   & 0.950  & 1.19             \\
                         & CE+Temp          & 0.901   & 0.950  & 1.18       \\
                         & CE+RT4U  & \textbf{0.907}   & 0.951  & 1.16            \\
                         & CE+RT4U+Temp  & \textbf{0.907}   & 0.951  & \textbf{1.14}   \\ \hline
\multirow{4}{*}{\begin{tabular}[c]{@{}l@{}l@{}}AS\\ Private \\ $\alpha = 0.1$ \end{tabular}} 
                         & R(2+1)D \cite{AS_Tom} & 0.780 & 0.913 & \textbf{1.35} \\                        
                         & R(2+1)D+RT4U      & \textbf{0.816}   & 0.917  & 1.36      \\
                         & ProtoASNet \cite{vaseli2023protoasnet}       & 0.789   & 0.924  & 1.40    \\
                         & ProtoASNet+RT4U   & 0.796   & 0.924  & 1.38  \\ \hline
\multirow{2}{*}{\begin{tabular}[c]{@{}l@{}}TMED-2\\ $\alpha = 0.1$ \end{tabular}}  & Huang et al. \cite{huang2021tmed1}            & 0.733   & 0.917  & \textbf{1.90}  \\
                         & Huang et al.+RT4U        & \textbf{0.742}   & 0.913  & 1.99  \\ \hline
\end{tabular}
\label{tab:study_level}
\end{table}

\begin{figure}[h]
    \centering
    \includegraphics[width=0.99\textwidth]{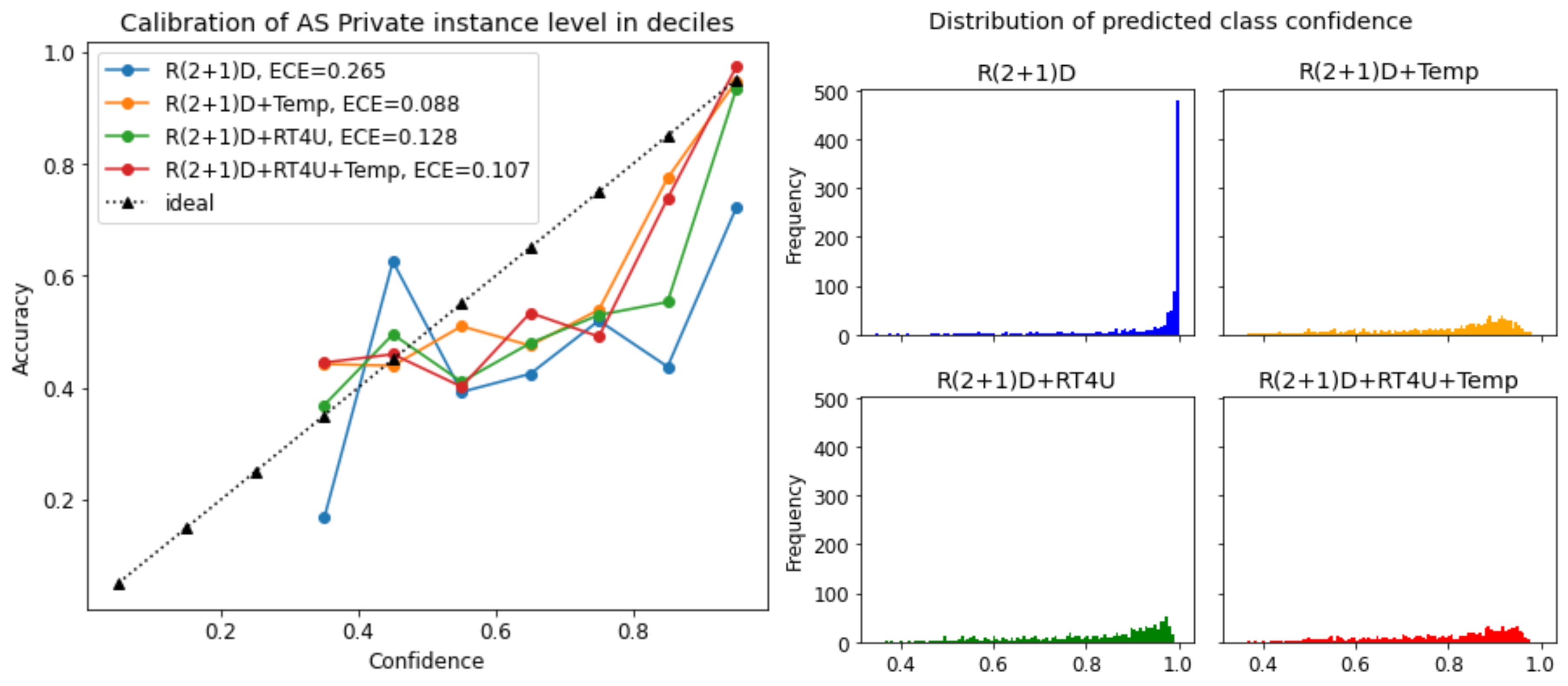}
    \caption{Left: Calibration performance of R(2+1)D variants with RT4U and temperature scaling, evaluated at the instance-level on the AS Private dataset. 
    Right: Histograms of network confidence for the predicted class. Compared to the baseline (blue), RT4U (green) has a lower expected calibration error, and the confidence distribution is closer to those from models with temperature scaling.
    }
    \label{fig:calibration}
\end{figure}

\subsection{Results and discussion}
Incorporating RT4U raises the $BAcc$ for each of the three AS severity classification methods (Tables \ref{tab:instance_level} and \ref{tab:study_level}). 
This is because RT4U reduces the loss induced by less informative inputs (for example, the PSAX view image with poor AVL visualization in Fig. \ref{fig:evolution}). 
In contrast, training with one-hot labels leads the network to eventually overfit to samples which do not contain any class-revealing information.
In both highly- and weakly- informative cases, RT4U sets the target to a more appropriate value. 

Training with one-hot labels results in a high degree of over-confidence (Fig. \ref{fig:calibration}). Conversely, training with RT4U makes the model slightly under-confident. This is because in RT4U the labels from highly-informative cases are also slightly smoothed to some value below $1.0$. Additionally, we would like to emphasize that prior to temperature scaling, the model trained with RT4U has a lower expected calibration error (ECE). Secondly, both methods can use temperature scaling in post-processing to reduce ECE.

The noise-robust loss functions, MAE and ANL \cite{ye2024anl}, under-perform relative to the cross-entropy baseline. This suggests that these methods, which are designed to handle noise from ``label flipping'', are susceptible to noise originating from the input.
Furthermore, the model trained with ANL produces uncalibrated outputs, which compromises the construction of prediction sets. 

Regarding the balanced coverage, our experiments confirm that the marginal coverage property is satisfied. 
The coverage is more conservative (i.e. a higher difference between $BCov$ and $1 - \alpha$) for AS Private and TMED-2, due to the smaller number of datapoints in their respective calibration sets. 
Furthermore, for disease classification, it is preferable for prediction sets to satisfy ``ordinality'', i.e. for any given input, there are no discontinuations within the predicted range of severities. We found that for every method tested on the TMED-2 and AS Private datasets, more than $98\%$ of prediction sets satisfy ordinality.

\section{Conclusion and Future Work}
We introduce RT4U, a novel training method aimed at tackling the issue of limited information within medical images. 
Firstly, the pseudo-labels provided by RT4U can mitigate overfitting to noisy inputs, and allow the network to associate these inputs to a high-uncertainty state. 
Secondly, in terms of conveying uncertainty, conformal prediction sets with a guaranteed accuracy offer a promising alternative to top-1 predictions and confidence values. Future improvements include introducing hyperparameters to refine the pseudolabel generation process, exploring the relationship between output logits and the size of prediction sets, and designing models with the primary objective of generating informative prediction sets rather than solely focusing on high top-1 accuracy.

\section{Acknowledgements}
This work was supported in part by the UBC Advanced Research Computing Center (ARC), the Canadian Institutes of Health Research (CIHR), and the Natural Sciences and Engineering Research Council of Canada (NSERC).

This preprint has not undergone any post-submission improvements or corrections. The Version of Record of this contribution is published in: International Conference on Medical Image Computing and Computer-Assisted Intervention (MICCAI), Springer (2024) under the same title.

\section{Disclosure of Interests}
The authors have no competing interests to declare that are
relevant to the content of this article.

%
%
%
\bibliographystyle{splncs04}
\bibliography{paper-2346}

\end{document}